\documentclass[11pt]{article}
\pdfoutput=1

\usepackage{jheppub}
\usepackage{bigints}

\usepackage{bbm}

\usepackage{bbold}

\usepackage{tabularx}

\usepackage{tikz}

\usepackage{ae}
\usepackage[T1]{fontenc}
\usepackage{url}

\usepackage{tabularx}

\usepackage{graphicx}
\usepackage{dsfont}
\usepackage{setspace}
\usepackage{amsfonts,amsmath,amsthm,amssymb,amsbsy}
\usepackage{longtable}
\usepackage[latin1]{inputenc}
\usepackage{array}
\usepackage{color}
\usepackage{needspace}
\usepackage[numbers]{natbib}
\usepackage{colortbl}

\usepackage{enumitem}

\usepackage{fancyhdr}

\title{The Momentum Amplituhedron}

\author[1]{David Damgaard,}\emailAdd{d.damgaard@lmu.de}
\author[1,2]{Livia Ferro,}\emailAdd{livia.ferro@lmu.de}
\author[2]{Tomasz \L ukowski,}\emailAdd{t.lukowski@herts.ac.uk}
\author[3]{and Matteo Parisi}\emailAdd{parisi@maths.ox.ac.uk}

\affiliation[1]{Arnold--Sommerfeld--Center for Theoretical Physics,\\ Ludwig--Maximilians--Universit\"at, \\ Theresienstra\ss e 37, 80333 M\"unchen, Germany }
\affiliation[2]{School of Physics, Astronomy and Mathematics, \\ University of Hertfordshire, \\  Hatfield, Hertfordshire, AL10 9AB, United Kingdom}
\affiliation[3]{Mathematical Institute, University of Oxford,\\ Andrew Wiles Building, Radcliffe Observatory Quarter,\\ Woodstock Road, Oxford, OX2 6GG, U.K.}

\abstract{In this paper we define a new object, the momentum amplituhedron, which is the long sought-after positive geometry for tree-level scattering amplitudes in $\mathcal{N}=4$ super Yang-Mills theory in spinor helicity space. Inspired by the construction of the ordinary amplituhedron, we introduce
 bosonized spinor helicity variables to represent our external kinematical data, and restrict them to a particular positive region.  The momentum amplituhedron $\mathcal{M}_{n,k}$ is then the image of the positive Grassmannian via a map determined by such kinematics.
  The scattering amplitudes are extracted from the canonical form with logarithmic singularities on the boundaries of this geometry. }

\begin{document}
\begin{flushright}
{\small LMU-ASC 21/19}
\end{flushright}
\maketitle


\section{Introduction}
In the past decade a new, geometric picture has emerged for scattering amplitudes in planar $\mathcal{N}=4$ super Yang-Mills (SYM) theory. It originated from the observation that the tree-level amplitudes and loop-level integrands of $n$-point amplitudes for all helicity sectors can be computed using  integrals over  the Grassmannian space \cite{ArkaniHamed:2009dn,Mason:2009qx}. In such formulation, amplitudes can be extracted from a Grassmannian integral over a suitable contour which selects a particular sum of residues. Building upon this idea, novel studies revealed the interrelation between the rich combinatorial structure of positive Grassmannians and the physical properties of amplitudes \cite{ArkaniHamed:2012nw}. From this point of view, the aforementioned residues are associated with positroid cells, which are particular subvarieties inside the positive Grassmannian. The proper combination of cells is selected by using the Britto-Cachazo-Feng-Witten (BCFW) recursion relations \cite{Britto:2004ap,Britto:2005fq}. However, the cells contributing to a particular amplitude are  seemingly not related to each other inside the positive Grassmannian. Nevertheless,  
via a map defined by a positive matrix of bosonized momentum twistor variables, they assemble in a convex-like object.  The image of the positive Grassmannian through such map is a geometric space, the {\it amplituhedron} \cite{Arkani-Hamed:2013jha}, and the union of the cell images provides a particular triangulation. 
The amplituhedron became eventually the first example of a vast family of the so-called {\it positive geometries} \cite{Arkani-Hamed:2017tmz}, which nowadays provide a geometric description for various quantities in theoretical physics: see, for instance, the kinematic associahedron  \cite{Arkani-Hamed:2017mur}, the cosmological polytope \cite{Arkani-Hamed:2017fdk}, and positive geometries in CFT \cite{Eden:2017fow,Arkani-Hamed:2018ign}.

Nevertheless, despite the name, the \emph{amplitu}hedron is more naturally suited to describe the dual Wilson loop rather than the amplitude itself, being defined in the momentum twistor space.  In particular, the employment of these variables restricts the possible generalization of this geometry to scattering amplitudes in other models, since it is based on the Amplitude/Wilson loop duality which is present only in planar $\mathcal{N}=4$ SYM. 
Therefore it limits the possibility of finding positive geometries for scattering amplitudes in less supersymmetric models and beyond the planar sector.
 It is then desirable to find a geometric description directly in the ordinary twistor space or, even better, in the spinor helicity space $(\lambda_i, \tilde\lambda_i)$. 
 The first attempt in this direction was made in \cite{He:2018okq}, where it was suggested that the amplituhedron in momentum space  should be the image of the twistor-string worldsheet \cite{Witten:2003nn} through the Roiban-Spradlin-Volovich (RSV) equations \cite{Roiban:2004yf}. In particular, it was conjectured that the space should have  proper sign flips for both $\lambda_i$ and $\tilde\lambda_i$, as well as the additional assumption that the planar Mandelstam variables should be positive. In this paper we will show that a suitable positive geometry with such characteristics exists and provides the proper expressions for the amplitude when written in the non-chiral superspace $(\lambda_i,\tilde\lambda_i,\eta_i,\tilde\eta_i)$. 
In order to achieve our goal, we will first introduce its bosonized version: $(\Lambda_i, \tilde{\Lambda}_i)$. By assuming that the external kinematic data $\Lambda$ and $\tilde\Lambda$ satisfy particular positivity conditions, we will reproduce the sign flips postulated in \cite{He:2018okq}. Additionally,  further constraints entangling $\Lambda$ and $\tilde\Lambda$ will enforce positivity of Mandelstam variables. Then, we will define the {\it momentum amplituhedron} as the image of the positive Grassmannian through a map determined by this positive external data. We will show that such space is indeed a positive geometry, whose canonical logarithmic differential form encodes scattering amplitudes in spinor helicity variables.

The paper is structured as follows. We start in section \ref{section.2} by reviewing the formulation of the original amplituhedron in the bosonized momentum twistor space. We proceed by defining the momentum amplituhedron, \emph{i.e.}  the positive geometry in the bozonized spinor helicity variables. Afterwards, we show how to find the logarithmic differential form on the momentum amplituhedron and how to extract the scattering amplitudes from it. Section \ref{section.3} consists of examples which show in detail  how to use the construction from section \ref{section.2}. We end the paper with Conclusions and Outlook, and a few Appendices containing more technical details of our construction.


\section{The Definition}\label{section.2}
\subsection{The Ordinary Amplituhedron}
We start by recalling the construction of the amplituhedron in momentum twistor space. In the past few years there has been a lot of progress on different descriptions of the amplituhedron \cite{Arkani-Hamed:2017vfh,Arkani-Hamed:2017tmz}. We will focus here on two of them, which will be relevant for our construction of the momentum amplituhedron: the original definition introduced in \cite{Arkani-Hamed:2013jha} and the description based on the sign flips presented in \cite{Arkani-Hamed:2017vfh}. The first states that the amplituhedron $\mathcal{A}_{n,k'}^{(m=4)}$ can be described on the space of bosonized supertwistors $Z^A_i$, $A=1,\ldots,4+k'$, which specify the kinematic data for the $n$-particle $\text{N}^{k'}\text{MHV}$ amplitude. The components of bosonized supertwistors include the bosonic part of the momentum supertwistors $(\lambda^{a}_i,\tilde\mu_i^{\dot a})$, $a,\dot a =1,2$, and the bosonized version of the fermionic components $\xi_i^\alpha=\phi^\alpha_\mathcal{R} \chi_i^\mathcal{R}$, $\alpha=1,\ldots k'$, where $\phi_\mathcal{R}^\alpha$ are auxiliary Grassmann-odd parameters and $\mathcal{R}=1,\ldots,4$ is the R-symmetry index. As already explored in the literature, there exists a straightforward generalization of bosonized variables beyond the case relevant for physics, $m=4$, and in the following we will allow any values for the label $m$. We start by demanding that the matrix of bosonized variables $Z=(Z_i^A)\in M_+(m+k',n)$ is positive, {\em i.e.} all its ordered maximal minors are positive. Then the amplituhedron $\mathcal{A}_{n,k'}^{(m)}$ is defined as the image of the map
\begin{equation}\label{PhiZ}
\Phi_Z:G_{+}(k',n)\to G(k',k'+m)\,,
\end{equation}
given by
\begin{equation}
Y^A_\alpha=c_{\alpha i}Z^A_i\in G(k',k'+m)\,, \qquad C=(c_{\alpha i})\in G_{+}(k',n)\,.
\end{equation}
Here, $G_{+}(k',n)$ is the positive Grassmannian, {\emph{i.e.}} the space of all positive matrices modulo $GL(k')$ transformations. For each amplituhedron $\mathcal{A}_{n,k'}^{(m)}$, 
one can define a $(k'\cdot m)$-dimensional differential form $\mathbf{\Omega}_{n,k'}^{(m)}$, called the {\em volume form}, which has logarithmic singularities on all boundaries (of all dimensions) of the amplituhedron space $\mathcal{A}_{n,k'}^{(m)}$. In particular, the volume form $\mathbf{\Omega}_{n,k'}^{(m=4)}$ encodes the N$^{k'}$MHV tree-level amplitude in $\mathcal{N}=4$ SYM.  The geometric space $\mathcal{A}_{n,k'}^{(m)}$ together with the form $\mathbf{\Omega}_{n,k'}^{(m)}$ describe a positive geometry, as defined in \cite{Arkani-Hamed:2017tmz}. Throughout the years, various methods to find the volume form $\mathbf{\Omega}_{n,k'}^{(m)}$ have been proposed \cite{Arkani-Hamed:2014dca,Ferro:2015grk,Arkani-Hamed:2017tmz,Ferro:2018vpf}. One can, for example, triangulate the amplituhedron $\mathcal{A}_{n,k'}^{(m)}$ by  {\emph{e.g.}} finding a collection of positroid cells $\mathcal{T}=\{\Delta_\sigma\}$ of dimension $k'\cdot m$ in $G_{+}(k',n)$ such that the images of these cells through the function $\Phi_Z$ do not overlap and cover the amplituhedron. To each positroid cell one can associate a canonical form $\omega_\sigma$ with logarithmic singularities on all its boundaries, see \cite{ArkaniHamed:2012nw}. The volume form $\mathbf{\Omega}_{n,k'}^{(m)}$ is found by evaluating the push-forward of the canonical forms $\omega_\sigma$ via the function $\Phi_Z$ and then summing over all positroid cells in the triangulation
\begin{equation}
\mathbf{\Omega}_{n,k'}^{(m)}=\sum_{\Delta_\sigma\in \mathcal{T}}(\Phi_Z)_*\,\omega_\sigma \,.
\end{equation}
The result of the push-forward is a logarithmic differential form on $G(k',k'+m)$ which can be written as
\begin{equation}\label{Omega.dlog}
\mathbf{\Omega}_{n,k'}^{(m)}=\sum_{\Delta_\sigma\in\mathcal{T}} \mathrm{d}_Y\mathrm{log}\,\alpha^\sigma_1(Y,Z)\wedge \mathrm{d}_Y\mathrm{log}\,\alpha^\sigma_2(Y,Z)\wedge \ldots\wedge \mathrm{d}_Y\mathrm{log}\,\alpha^\sigma_{k'm}(Y,Z)\,,
\end{equation}
where $\alpha^\sigma_i(Y,Z)$ are the canonical positive coordinates parametrizing the cell $\Delta_\sigma$. The explicit expressions for various values of parameters $(n,m,k')$ can be found e.~g.~in \cite{Arkani-Hamed:2017tmz}.

An alternative way to find the volume form is to introduce a {\em volume function} $\Omega_{n,k'}^{(m)}$ defined by
\begin{equation}
\mathbf{\Omega}_{n,k'}^{(m)}=\prod_{\alpha=1}^{k'} \langle Y_1\ldots Y_{k'}\mathrm{d}^m Y_\alpha \rangle \, \Omega_{n,k'}^{(m)}\,,
\end{equation}
where $\prod_{\alpha=1}^{k'} \langle Y_1\ldots Y_{k'}\mathrm{d}^m Y_\alpha \rangle$ is the standard measure on the Grassmannian $G(k',k'+m)$.
The volume function can also be  obtained  by evaluating the integral over the space of $k\times n$ matrices
\begin{equation}
\Omega_{n,k'}^{(m)}=\int_\gamma \frac{\mathrm{d}^{k'\cdot n}c_{\alpha i}}{(12\ldots k')(23\ldots k'+1)\ldots (n1\ldots k'-1)}\prod_{\alpha=1}^{k'}\delta^{m+k'}(Y^A_\alpha-\sum_i c_{\alpha i}Z^A_i)\,,
\end{equation}
over a suitable contour $\gamma$. The integrand is a meromorphic function of $c$ and the integral reduces to a sum of residues, specified by the contour $\gamma$.

The original construction of the amplituhedron defines the volume form $\mathbf{\Omega}_{n,k'}^{(m)}$ as differential form on an auxiliary Grassmannian space $G(k',k'+m)$ parametrized by $Y$. However, as pointed out in \cite{Arkani-Hamed:2017vfh}, the form \eqref{Omega.dlog} can be also thought of as a differential form on the purely bosonic part of the momentum supertwistors, $z_i^a$, $a=1,\ldots,m$. It can be accomplished by replacing the differential with respect to $Y$ by the differential with respect to the kinematic data $Z$: $\mathrm{d}_Y\mathrm{log} \to \mathrm{d}_Z \mathrm{log}$, and at the same time by fixing $Y=Y^*$, where $Y^*$ is a reference $k'$-plane in $k'+m$ dimensions. This new differential form is a logarithmic differential form on the space of configurations of  $z_i^a$, satisfying particular sign-flip or topological conditions \cite{Arkani-Hamed:2017vfh}. Let us now recall the proper sign-flip conditions in the $m=2$ case, which will be relevant for us in the following. For $m=2$, we consider a configuration of two-dimensional vectors $z_i^a$, $a=1,2$, and define the brackets $\langle ij\rangle_z:=z_i^1 z_j^2-z_i^2 z_j^1$. The amplituhedron $\mathcal{A}_{n,k'}^{(m=2)}$ space is defined as a subspace of the configuration space $\{z_i\}_{i=1,\ldots,n}$ satisfying the following conditions: 
\begin{equation}
\langle i i+1\rangle_z >0 \,\text{and the sequence} \left\{ \langle 12\rangle_z,\langle 13\rangle_z,\ldots ,\langle 1n\rangle_z \right\} \text{has exactly $k'$ sign flips.} 
\end{equation}
Although the sign-flip characterization of the amplituhedron does not refer either to any auxiliary space or the quite peculiar bosonization described above, it is not an easy task to find the volume form directly from this definition. Therefore, we often refer back to the original construction of the amplituhedron in the bosonized space. 

\subsection{The Momentum Amplituhedron}
In order to define an amplituhedron directly in the spinor helicity space we will follow a reverse path compared to the one described in the previous section. Our starting point will be the conjecture in \cite{He:2018okq} suggesting that we should consider a positive geometry described by proper sign-flips in the spinor helicity space, together with  positivity of the Mandelstam variables formed out of consecutive momenta. Let us start by taking a configuration space of $n$ spinor-helicity variables parametrized by $\{\lambda^a,\tilde\lambda^{\dot a}\}$, $a,\dot a=1,2$, and define the brackets $\langle ij\rangle_\lambda:=\lambda_i^1\lambda^2_j-\lambda_i^2\lambda^1_j$ and $[ ij]_{\tilde\lambda}:=\tilde\lambda_i^1\tilde\lambda^2_j-\tilde\lambda_i^2\tilde\lambda^1_j$. Let us also define {\em planar Mandelstam variables}
\begin{equation}\label{Mandelstams}
s_{i,i+1,\ldots,i+p}=\sum_{i\leq j_1<j_2\leq i+p}\langle j_1 j_2\rangle_\lambda [j_1j_2]_{\tilde\lambda}\,.
\end{equation} 
This relation is understood modulo $n$.
Then the conjecture of \cite{He:2018okq} states that the positive region would be defined by the following conditions:
\begin{itemize}
\item Positive planar Mandelstam variables: $s_{i,i+1,\ldots,i+p}>0$ for $i=1,\ldots,n$, $p=1,\ldots,n-3$.
\item Correct sign flips: let the list $\{\langle12\rangle_{\lambda},\langle13\rangle_{\lambda},\ldots,\langle1n\rangle_{\lambda}\}$ have $N$ sign flips and the list $\{[12]_{\tilde\lambda},[13]_{\tilde\lambda},\ldots,[1n]_{\tilde\lambda}\}$ have $\tilde{N}$ sign flips, then we require one of the two possibilities: $(N,\tilde{N})=(k-2,k)$ or $(N,\tilde{N})=(n-k,n-k-2)$.
\end{itemize}
In this paper we define the space of bosonized spinor helicity variables and a related positive geometry, which we call the {\em momentum amplituhedron} $\mathcal{M}_{n,k}$. This will encode the N$^{k-2}$MHV $n$-particle tree-level amplitudes in $\mathcal{N}=4$ SYM\footnote{Notice that $k=k'+2$, where $k'$ was defined in the previous section.}. By demanding certain positivity conditions on the bosonized variables, we will recover proper sign flips for $\lambda$ and $\tilde\lambda$. With additional assumptions, we will also guarantee that the planar Mandelstam variables are positive.

As described in details in \cite{He:2018okq}, any $n$-particle N$^{k-2}$MHV  scattering amplitude in planar $\mathcal{N}=4$ SYM can be written as a differential form in spinor helicity space. The starting point is the non-chiral superspace which is parametrized by spinor helicity variables, $\{\lambda^a,\tilde\lambda^{\dot a}\}$, $a,\dot a=1,2$, together with the Grassmann odd parameters $\{\eta^r,\tilde\eta^{\dot r}\}$, $r,\dot r=1,2$. Then the amplitude is a function on $n$ copies of this superspace with coordinates $\{\lambda_i,\eta_i|\tilde\lambda_i,\tilde\eta_i\}$. Let us remark that in this space the supercharges take the form:
\begin{equation}
\label{qtq}
\tilde q^{\dot a r}=\sum_{i=1}^n\tilde\lambda_i^{\dot a}\eta_i^r \quad , \quad  q^{a r}=\sum_{i=1}^n\lambda_i^{a}\tilde\eta_i^{\dot r} \,.
\end{equation}
Moreover, there is a natural way to associate the R-symmetry indices $(r,\dot r)$ with the spinor indices $(a,\dot a)$ and write any function on the superspace as a differential form on its bosonic part. It amounts to the replacement
\begin{equation}
\label{lambda_eta}
\eta^a \to d\lambda^a\, ,\qquad\qquad \tilde\eta^{\dot a}\to d\tilde\lambda^{\dot a}\,.
\end{equation} 
The degree of this differential form in $(d\lambda,d\tilde\lambda)$ is then $(2(n-k),2k)$, respectively. This is a similar situation to the one which we have encountered for the momentum twistors, where the amplitude can be thought of as a differential form of degree $4k'=4(k-2)$ on the bosonic part of the momentum twistor superspace. Equivalently, it was possible to introduce a bosonized momentum twistor space by introducing auxiliary Grassmann-odd parameters. We will now repeat this construction for the spinor helicity variables.

Let us introduce $2(n-k)$ auxiliary Grassmann-odd parameters $\phi_{a}^\alpha$, $\alpha=1,\ldots,n-k$ and $2k$ auxiliary Grassmann-odd parameters $\tilde\phi_{\dot a}^{\dot\alpha}$, $\dot\alpha=1,\ldots,k$. We define bosonized spinor helicity variables as
\begin{align}
\label{bosonizedY}
&\Lambda^A_i=\left(\begin{tabular}{c}$\lambda^a_i$\\$\phi_{a}^\alpha\cdot\eta^a_i$\end{tabular}\right),\qquad A=(a,\alpha)=1,\ldots,n-k+2\,,\\ 
\label{bosonizedYt} &\tilde\Lambda^{\dot A}_i=\left(\begin{tabular}{c}$\tilde\lambda^{\dot a}_i$\\$\tilde\phi_{\dot a}^{\dot\alpha}\cdot\tilde\eta^{\dot a}_i$\end{tabular}\right),\qquad \dot{A}=(\dot a,\dot\alpha)=1,\ldots,k+2\,.
\end{align}
In the next step we define a positive region on the space of bosonized spinor helicity variables. We introduce the matrices
\begin{equation}
\Lambda=\left(\begin{matrix}
\Lambda_{1}^A&\Lambda_{2}^A&\ldots&\Lambda_n^A
\end{matrix}\right)\in M(n-k+2,n),\qquad \tilde\Lambda=\left(\begin{matrix}
\tilde\Lambda_{1}^{\dot{A}}&\tilde\Lambda_{2}^{\dot{A}}&\ldots&\tilde\Lambda_n^{\dot{A}}
\end{matrix}\right)\in M(k+2,n)\,,
\end{equation}
and refer to the pair $(\Lambda,\tilde\Lambda)$ as the {\em kinematic data}.
These matrices describe linear subspaces of dimension $n-k+2$ and $k+2$, respectively, inside an $n$-dimensional space. We denote their orthogonal complements as $\Lambda^\perp\in M(k-2,n)$ and $\tilde\Lambda^\perp\in M(n-k-2,n)$. The orthogonal complements are defined up to a $GL$-transformation, corresponding to a change of basis of the corresponding subspaces.
Additionally, we define two types of brackets on the space of bosonized variables. On the space of $\Lambda$'s we define
\begin{equation}
\langle i_1 i_2\ldots i_{n-k+2}\rangle=\epsilon_{A_1A_2\ldots A_{n-k+2}}\Lambda_{i_1}^{A_1}\Lambda_{i_2}^{A_2}\ldots  \Lambda_{i_{n-k+2}}^{A_{n-k+2}}\,.
\end{equation}
Similarly for the space of $\tilde\Lambda$'s we have
\begin{equation}
[i_1 i_2\ldots i_{k+2}]=\epsilon_{\dot A_1\dot A_2\ldots \dot A_{k+2}}\tilde\Lambda_{i_1}^{\dot A_1}\tilde\Lambda_{i_2}^{\dot A_2}\ldots  \tilde\Lambda_{i_{k+2}}^{\dot A_{k+2}}\,.
\end{equation}

Until now there is manifest symmetry between $\Lambda$ and $\tilde\Lambda$: if we exchange $\Lambda\leftrightarrow \tilde\Lambda$ together with exchanging $k\leftrightarrow n-k$, the space looks the same. This corresponds to the parity invariance of $\mathcal{N}=4$ SYM. In the following we will, however, need to break this symmetry by choosing one of two possible descriptions. These two choices correspond to the two possibilities available in the conjecture for the sign-flip condition. In order to define the {\em positive region}, we restrict the allowed external data to be positive in the following sense:
\begin{align}
 \left\{\begin{array}{l}
\text{matrix} ~ \tilde\Lambda ~ \text{positive}  \\
\text{matrix}~ \Lambda^\perp ~\text{positive}
\end{array}
\right\} \,.
\end{align}
Alternatively, we could assume that the matrices $\tilde\Lambda^\perp$ and $\Lambda$ are positive and proceed in an analogous way.
We emphasize that the fact that the matrix $\Lambda^\perp$ is positive does not imply that the matrix $\Lambda$ is positive. On the contrary, using the discussion from the Appendix \ref{App:perp}, one can notice that the matrix encoding the orthogonal complement of a positive matrix will have both positive and negative minors.

Having defined the positive region we are ready to adapt the map \eqref{PhiZ} to the bosonized spinor helicity space. We define the momentum amplituhedron $\mathcal{M}_{n,k}$ as the image of the positive Grassmannian $G_+(k,n)$ through the map
\begin{equation}
\label{Phi}
\Phi_{(\Lambda,\tilde\Lambda)}:G_+(k,n)\to G(k,k+2)\times G(n-k,n-k+2)\,,
\end{equation}
which to each element of the positive Grassmannian $C=\{c_{\dot\alpha i}\}\in G_{+}(k,n)$ associates a pair of Grassmannian elements $(\tilde Y,Y)\in G(k,k+2)\times G(n-k,n-k+2)$ in the following way
\begin{align}
\tilde Y^{\dot{A}}_{\dot{\alpha}}=c_{\dot\alpha i}\,\tilde\Lambda_i^{\dot{A}}\,,\qquad\qquad Y^A_\alpha=c^\perp_{\alpha i}\,\Lambda_i^A\,,
\label{Y}
\end{align}
where $C^\perp=\{c^\perp_{\alpha i}\}$ is the orthogonal complement of $C$. One can show that $Y$ has rank $(n-k)$, therefore it is an element of $G(n-k,n-k+2)$ and the map $\Phi_{(\Lambda,\tilde\Lambda)}$ is well defined. After imposing additional assumptions on $\Lambda$ and $\tilde\Lambda$, which will guarantee positive planar Mandelstam variables, we claim that the momentum amplituhedron $\mathcal{M}_{n,k}$ is a positive geometry and its volume form encodes the $n$-particle N$^{k-2}$MHV tree-level scattering amplitude in $\mathcal{N}=4$ SYM.

In order to confirm this claim, we start by checking that the momentum amplituhedron has the expected dimension, namely $2n-4$, and that we find the correct pattern of sign flips. Let us first observe  that the dimension of $G(k,k+2)\times G(n-k,n-k+2)$ is $2n$
\begin{equation}
\dim(G(k,k+2))+\dim(G(n-k,n-k+2))=2k+2(n-k)=2n\,.
\end{equation}
We notice, however, that the image of the positive Grassmannian $G_{+}(k,n)$ through the map $\Phi_{(\Lambda,\tilde\Lambda)}$  is lower dimensional. Indeed,  the momentum amplituhedron lives in the following  co-dimension four surface inside $G(k,k+2)\times G(n-k,n-k+2)$:
\begin{equation}\label{momentum.cons}
P^{a\dot a}=\sum_{i=1}^n\left(Y^\perp \cdot \Lambda\right)^a_i \left(\tilde Y^\perp \cdot \tilde\Lambda \right)^{\dot a}_i=0\,.
\end{equation} 
For a proof of this statement see Appendix \ref{App.mom.cons}.
We defined here the orthogonal complements $Y^\perp\in G(2,n-k+2)$ and $\tilde Y^\perp\in G(2,k+2)$. One can think about the condition \eqref{momentum.cons} as being equivalent to the momentum conservation but written directly in the momentum amplituhedron space. Indeed, if we project through a fixed $Y$ and $\tilde Y$, as we will see later, then we find
\begin{equation}
\label{projectY}
\left(Y^\perp \cdot \Lambda\right)^a_i  \to \lambda_i^a\,,\qquad\qquad \left(\tilde Y^\perp \cdot \tilde\Lambda \right)^{\dot a}_i\to\tilde\lambda^{\dot a}_i\,,
\end{equation} 
and the condition \eqref{momentum.cons} reduces to the usual momentum conservation.
Equation \eqref{momentum.cons} implies that the image of the positive Grassmannian $G_+(k,n)$ through the map $\Phi_{(\Lambda,\tilde\Lambda)}$ is a co-dimension four surface inside the space $ G(k,k+2)\times G(n-k,n-k+2)$ and therefore has the correct dimension $2n-4$.

The second check we would like to perform is to confirm that this geometry satisfies the correct sign flip conditions, postulated in \cite{He:2018okq}. Let us first remind the reader that one can reduce the geometry in the bosonized space to the purely bosonic part by projecting the kinematic configuration in the direction of a fixed $Y$, see \cite{Arkani-Hamed:2017vfh}. In the context of the momentum amplituhedron, the projection results in the reduction:
\begin{align}
\langle Yij\rangle \to \langle ij\rangle_\lambda\,,\qquad\qquad[ \tilde{Y} ij] \to [ ij]_{\tilde\lambda}\,.
\end{align}
Therefore, we are interested in the following sequences of brackets:
\begin{equation}\label{Y.flips}
\{\langle Y12\rangle,\langle Y13\rangle,\ldots,\langle Y1n\rangle\}\,,
\end{equation}
and
\begin{equation}\label{Ytilde.flips}
\{[ \tilde Y12],[\tilde Y13],\ldots,[ \tilde Y1n]\}\,.
\end{equation}
We want to show that the number of sign flips  equals $k-2$ in the sequence \eqref{Y.flips} and $k$ in the sequence \eqref{Ytilde.flips}. This corresponds to the condition $(N,\tilde{N})=(k-2,k)$ in the conjecture in \cite{He:2018okq}.
It is easy to see that the number of sign flips in the sequence \eqref{Ytilde.flips} is $k$ since the formula in \eqref{Y} for $\tilde Y$ is the definition of the ordinary amplituhedron \cite{Arkani-Hamed:2013jha} with $m=2$ and $k=k'$. It was shown in \cite{Arkani-Hamed:2017vfh} that in this case the number of sign flips equals $k$. The sequence \eqref{Y.flips} requires further attention. Let us define $X \,\in G(k-2,k)$ by
\begin{equation}\label{X.def}
X^{\bar{A}}_{\dot\alpha}=(\Lambda^\perp)_i^{\bar{A}}\,c_{\dot\alpha i}\,, \quad \bar{A}=1,\ldots, k-2 \,, \, \dot\alpha=1,\ldots,k \,.
\end{equation}
We emphasize that both matrices $\Lambda^\perp$ and $C$ in \eqref{X.def} are positive. Therefore \eqref{X.def} is similar to the definition of the ordinary amplituhedron with $m=2$ and $k\to k-2$, with the role of the matrices $C$ and $\Lambda^\perp$ exchanged. It implies that the number of sign flips in the sequence
\begin{equation}
\{(X12),(X13),\ldots,(X1n)\}\,,
\end{equation}
equals $k-2$, where we defined
\begin{equation}
(Xij) = \epsilon_{\dot\alpha_1\ldots \dot\alpha_k}X^1_{\dot\alpha_1}....X^{k-2}_{\dot\alpha_{k-2}} c_{\dot\alpha_{k-1},i} \,c_{\dot\alpha_k,j} \,.
\end{equation}
Moreover, one can show that 
\begin{equation}
\label{XandY}
(Xij)=\langle Yij\rangle\,,
\end{equation}
see Appendix \ref{App:XandY}. This implies that the number of sign flips in \eqref{Y.flips} is $k-2$, as required.

\subsection{Momentum Amplituhedron Volume Form}\label{Sec:boundaries}
Having defined the space $\mathcal{M}_{n,k}$, we want to find its volume form, {\it i.e.}~the differential form with logarithmic singularities on all boundaries of $\mathcal{M}_{n,k}$. We start by classifying possible boundaries of the momentum amplituhedron. There are three different types of boundaries:  two of them are similar to the ones we have encountered already for the amplituhedron $\mathcal{A}_{n,k}^{(m=2)}$:
\begin{equation}\label{boundaries.two}
\langle Y \,i \,i+1\rangle=0\,, \qquad\qquad [\tilde Y\,i\,i+1]=0\,.
\end{equation}
These can be related to all possible collinear limits of the amplitude.
In addition, there is also a new type of boundary which depends on both $\Lambda$ and $\tilde\Lambda$. These are defined by 
\begin{equation}\label{boundaries.mandelstam}
S_{i,i+1\ldots,i+p}=0\,,\qquad p=2,\ldots,n-4\,,
\end{equation} 
where $S_{i,i+1\ldots,i+p}$ is the uplift of the planar Mandelstam variables  \eqref{Mandelstams}  to the amplituhedron space defined as
\begin{eqnarray}
S_{i,i+1\ldots,i+p} &=&  \sum_{\substack{i\leq j_1<j_2\leq i+p}} \langle Y j_1 j_2 \rangle [\tilde{Y}j_1 j_2] \,.
\end{eqnarray}
Notice that $S_{i,i+1\ldots,i+p}$ reduces to the ordinary Mandelstam variables $s_{i,i+1\ldots,i+p} $ we defined in \eqref{Mandelstams}  when projected through fixed $Y$ and $\tilde Y$. The boundaries \eqref{boundaries.mandelstam} correspond to all possible non-trivial factorizations of the amplitude. Notice that the case when a two-particle Mandelstam variable vanishes splits into two boundaries of the momentum amplituhedron of the type \eqref{boundaries.two} and are not included in \eqref{boundaries.mandelstam}.

We look now for a differential form $\mathbf{\Omega}_{n,k}$ with logarithmic singularities on all boundaries of the form \eqref{boundaries.two} and \eqref{boundaries.mandelstam} and which is finite inside $\mathcal{M}_{n,k}$.
To do this we first triangulate the space $\mathcal{M}_{n,k}$ with each triangle being an image through the map $\Phi_{(\Lambda,\tilde\Lambda)}$ of a $(2n-4)$-dimensional cell of the positive Grassmannian $G_{+}(k,n)$. The proper combination of cells can be found using the {\tt positroid} Mathematica\texttrademark~package \cite{Bourjaily:2012gy}\footnote{To find a possible triangulation of $\mathcal{M}_{n,k}$ one needs to use the function {\texttt treeContour[n,k]}.}.  The logarithmic differential form on $\mathcal{M}_{n,k}$ is the sum over such cells of push-forwards of canonical differential form for each cell. As for the ordinary amplituhedron $\mathcal{A}_{n,k}^{(m)}$, the explicit answer is a sum of rational functions where the denominators can contain spurious singularities, corresponding to spurious boundaries in a given triangulation. These singularities disappear in the complete sum and the only divergences of $\mathbf{\Omega}_{n,k}$ correspond to the external boundaries \eqref{boundaries.two} and \eqref{boundaries.mandelstam}. The final check we need to perform in order to obtain a positive geometry is to confirm that there are no singularities of $\mathbf{\Omega}_{n,k}$ inside $\mathcal{M}_{n,k}$. It is clear for the   boundaries \eqref{boundaries.two} because it is easy to show that for all points inside the amplituhedron
\begin{equation}
\langle Y \,i \,i+1\rangle>0\,, \qquad\qquad [\tilde Y\,i\,i+1]>0\,.
\end{equation}
The situation is more complicated for the singularities where $S_{i,i+1,\ldots,i+p}$ vanish. As we will see when studying examples in the following section, the positivity conditions which we spelled out in the previous section -- $\tilde\Lambda$ positive and $\Lambda^{\perp}$ positive -- are not enough to guarantee that $S_{i,i+1,\ldots,i+p}>0$ for all points inside the amplituhedron $\mathcal{M}_{n,k}$. At the moment it is unclear in full generality what are the necessary and sufficient conditions to enforce positive Mandelstam variables. Nevertheless, we have found instances for which all planar Mandelstams are positive for all points in $\mathcal{M}_{n,k}$, proving that the set of configurations for which the momentum amplituhedron is a positive geometry is non-empty. Let us take for example the following parametrization of the kinematic data:
\begin{equation}\label{moment.curve}
(\Lambda^\perp)_i^{\bar{A}}=i^{\bar{A}-1}\,,\qquad \tilde\Lambda_i^{\dot A}=i^{\dot A-1}\,.
\end{equation}
This choice of positive matrices corresponds to considering the vertices of the polytopes defined by the matrices $\Lambda^\perp$ and $\tilde\Lambda$  to lie on the moment curve. 
We have explicitly checked that for all points inside the momentum amplituhedron $\mathcal{M}_{n,k}$, for $n\leq 10$ and any $k$, and with the kinematic data specified by \eqref{moment.curve}, all planar Mandelstam variables are positive:
\begin{equation}
S_{i,i+1\ldots,i+p}>0\,,\qquad p=1,\ldots,n-3\,.
\end{equation}  
We will study examples in more detail in the following section. We will notice that the space of allowed kinematic configurations is rather large and, in particular, for MHV and $\overline{\text{MHV}}$ amplitudes all kinematic configurations provide positive geometry.

We conclude this section by two remarks. First, we describe how to obtain the amplitude $\mathcal{A}^{\tt tree}_{n,k}$ from the volume form $\mathbf{\Omega}_{n,k}$.  Let us recall that the momentum amplituhedron $\mathcal{M}_{n,k}$ is $(2n-4)$-dimensional and therefore the degree of $\mathbf{\Omega}_{n,k}$ is $(2n-4)$. There are various ways one can write $\mathbf{\Omega}_{n,k}$, related to each other by momentum conservation. In order to make it invariant we use the fact that $1=\delta^4(P) d^4P$. This allows us to define the volume function $\Omega_{n,k}$ in the following way:
\begin{equation}
\label{volumefc}
\mathbf{\Omega}_{n,k}  \wedge d^4P \, \delta^4(P) =\prod_{\alpha=1}^{n-k} \langle Y_1\ldots Y_{n-k} d^2 Y_\alpha \rangle \prod_{\dot{\alpha}=1}^{k} [ \tilde Y_1\ldots \tilde Y_{k} d^2 \tilde Y_{\dot{\alpha}}]   \, \delta^4(P)\,\Omega_{n,k} \,.
\end{equation}
Indeed, the form $\mathbf{\Omega}_{n,k}  \wedge d^4P$ is top-dimensional and therefore can be written in terms of the measure on $G(k,k+2) \times G(n-k,n-k+2)$ multiplied by a function.
Then, the procedure to extract the amplitude from the volume form $\mathbf{\Omega}_{n,k}$ is similar to the ordinary amplituhedron, {\it i.e.}~we localize the $Y$ and $\tilde Y$ on reference subspaces\footnote{This choice of $Y^*, \tilde Y^*$ is compatible with the embedding of  $\lambda,\tilde \lambda$ in  $\Lambda, \tilde\Lambda$ as in \eqref{bosonizedY}, \eqref{bosonizedYt}.}
 \begin{equation}
 Y^*=\left(\begin{matrix}
\mathbb{0}_{2\times (n-k)}\\
\hline
\mathbb{1}_{(n-k)\times (n-k)}
\end{matrix}\right),\qquad\qquad 
\tilde Y^*=\left(\begin{matrix}
\mathbb{0}_{2\times k}\\
\hline
\mathbb{1}_{k\times k}
\end{matrix}\right) ,
\end{equation}  
obtaining
 \begin{equation}
 \label{extract}
 \mathcal{A}^{\tt tree}_{n,k}=   \delta^4(p) \int d \phi^1_a \ldots  d\phi^{n-k}_a\int d\tilde\phi^1_{\dot a}\ldots  d\tilde\phi^k_{\dot a}\, \,\Omega_{n,k}(Y^*,\tilde Y^*,\Lambda,\tilde\Lambda) \,,
 \end{equation}
 where $\delta^4(p)$ comes from the localization of $\delta^4(P)$ on $Y^*,\tilde Y^*$.
 In the following section we will show how extracting the amplitude works in practice in a few examples.

Finally, in analogy with the ordinary amplituhedron, we can introduce an integral representation of the volume function $\Omega_{n,k}$ as an integral over a matrix space
 \begin{equation}
\delta^4(P) \,\Omega_{n,k} = \int 
\frac{\mbox{d}^{(n-k)\cdot (n-k)}{g}}{(\mbox{det}{g})^{n-k}}  \, \int \omega_{n,k}  \prod_{\alpha=1}^{n-k} \delta^{(n-k+2)}(Y^A_\alpha-g_{\alpha}^{\beta}\,(c^\perp)_{\beta i}\,\Lambda_i^A)  \prod_{\dot\alpha=1}^{k} \delta^{(k+2)}(\tilde Y^{\dot{A}}_{\dot{\alpha}}-\,c_{\dot\alpha i}\,\tilde\Lambda_i^{\dot{A}})\,,
\end{equation}
where we additionally need to integrate over the matrix $g$ corresponding to a $GL(n-k)$-transformation encoding the ambiguity of defining an orthogonal complement. The integration measure $\omega_{n,k}$ is the canonical measure on the space of $k\cdot n$ matrices:
\begin{equation}
\omega_{n,k} = \frac{d^{k\cdot n}c_{\dot\alpha i}}{(12\ldots k)(23\ldots k+1)\ldots (n1\ldots k-1)}\,,
 \end{equation}
 where the brackets in the denominator are minors of the matrix $C$
\begin{equation}\label{c.minors}
(i_1 i_2\ldots i_k)=\epsilon_{\dot\alpha_1\dot\alpha_2\ldots\dot\alpha_k}\,c_{\dot\alpha_1i_1}c_{\dot\alpha_2i_2}\ldots c_{\dot\alpha_ki_k}\,.
\end{equation}


\section{Examples}
\label{section.3}

\subsection{MHV/$\overline{\text{MHV}}$ Amplitudes}

We now move to study examples of momentum amplituhedra, starting with MHV and $\overline{\text{MHV}}$ amplitudes. Already in this case the volume function takes a new and interesting form. The dimension of the momentum amplituhedron $\mathcal{M}_{n,2}$ is the same as the dimension of the positive Grassmannian $G_+(2,n)$ and therefore there is no need to triangulate the amplituhedron, it is enough to take the image of the Grassmannian top-dimensional positroid cell. It is an easy task to find all boundaries of the momentum amplituhedron $\mathcal{M}_{n,2}$: they are all of the form $\langle Y i i+1\rangle =0$ for $i=1,\ldots,n$. The volume form we find in this section will make these boundaries manifest. One can also show that, for all points inside the momentum amplituhedron $\mathcal{M}_{n,2}$,  $[\tilde Y ii+1]>0$ for all $i=1,\ldots,n$, as well as $S_{i,i+1\ldots, i+p}>0$ for all $i=1,\ldots ,n$ and $p=1,\ldots,n-3$, see appendix \ref{app:positiveM}.

Let us start by considering the simplest case, {\it i.e.}~the four-point MHV amplitude.  We parametrize the top cell of $G_{+}(2,4)$ using the positive parameters $\alpha_j$:
\begin{equation}
C=\left( \begin{matrix}
1&\alpha_2&0&-\alpha_3\\
0&\alpha_1&1&\alpha_4
\end{matrix}\right) \,.
\end{equation} 
There are various ways to find $\alpha$'s from equations \eqref{Y}. A particular choice results in $\alpha$'s depending only on $Y$ and $\Lambda$:
\begin{equation}
\alpha_1=\frac{\langle Y12\rangle}{\langle Y13\rangle} \,,\alpha_2=\frac{\langle Y23\rangle}{\langle Y13\rangle} \,, \alpha_3=\frac{\langle Y34\rangle}{\langle Y13\rangle}\,, \alpha_4=\frac{\langle Y14\rangle}{\langle Y13\rangle} \,.
\end{equation}
The push-forward of the Grassmannian top form through \eqref{Phi} is therefore:
\begin{align}\label{ansY.n4k2}
 \mathbf{\Omega}_{4,2}&=\bigwedge_{j=1}^4\mathrm{dlog}\alpha_j = \mathrm{dlog}\frac{\langle Y12\rangle}{\langle Y13\rangle} \wedge \mathrm{dlog}\frac{\langle Y23\rangle}{\langle Y13\rangle} \wedge \mathrm{dlog}\frac{\langle Y34\rangle}{\langle Y13\rangle} \wedge \mathrm{dlog}\frac{\langle Y14\rangle}{\langle Y13\rangle}  \\
&= \frac{\langle 1234\rangle^2}{\langle Y12\rangle \langle Y23\rangle \langle Y34\rangle \langle Y41\rangle} \langle Y d^2Y_1\rangle\langle Y d^2Y_2\rangle  \,.
\end{align}
If instead we solve equations \eqref{Y} only in terms of $\tilde Y$ we find the following representation for the volume form
\begin{align}
\label{ansYt.n4k2}
 \mathbf{\Omega}_{4,2} = \frac{[1234]^2}{[\tilde Y 12] [\tilde Y 23] [\tilde Y 34] [\tilde Y 41] } [ \tilde Y d^2\tilde Y_1][ \tilde Y d^2\tilde Y_2]  \,.
\end{align}
It is easy to check that \eqref{ansY.n4k2} and \eqref{ansYt.n4k2} are related to each other by momentum conservation \eqref{momentum.cons}. Independently of the chosen representation for the volume form, the volume function can be evaluated using \eqref{volumefc} and gives the following manifestly parity symmetric answer:
\begin{equation}
\Omega_{4,2} = \frac{\langle 1234\rangle^2 [1234]^2}{\langle Y12\rangle \langle Y23\rangle [\tilde Y 12] [\tilde Y 23]} \,,
\end{equation}
unique up to momentum conservation.
Finally, we can extract the amplitude $\mathcal{A}^{\tt tree}_{4,2}$ using \eqref{extract} to get
\begin{equation}
\mathcal{A}^{\tt tree}_{4,2}=\delta^4(p)\frac{\delta^4(q)\delta^4(\tilde q)}{\langle 12\rangle_\lambda\langle 23\rangle_\lambda [12]_{\tilde\lambda} [23]_{\tilde\lambda}}\,,
\end{equation}
where $q, \tilde q$ are defined in \eqref{qtq}. 
 This formula agrees with the result found in \cite{He:2018okq}.

This calculation can be easily generalized to any $\mathrm{MHV}$ amplitude.  A particular representation for the volume form reads
\begin{eqnarray}\label{Omegan2log}
\mathbf{\Omega}_{n,2} &=& \bigwedge_{i=2}^{n-1}\left(\mbox{dlog} \left( \frac{\langle Y i,i+1\rangle}{\langle Y 1, i+1 \rangle} \right) \wedge \mbox{dlog} \left( \frac{\langle Y 1,i+1\rangle}{\langle Y 12 \rangle} \right) \right)\\\label{Omegan2}
&=&  \frac{ \langle 1\dots n\rangle^2}{\langle Y12\rangle \langle Y23\rangle \dots \langle Y1n\rangle} \langle Y \mbox{d}^{2}Y_1 \rangle \langle Y \mbox{d}^{2}Y_2 \rangle \dots  \langle Y \mbox{d}^{2}Y_{n-2} \rangle 
 \,.
\end{eqnarray} 
This result agrees with the one we get for the ordinary amplituhedron $\mathcal{A}_{n,n-2}^{(2)}$. Let us notice that, when $\mathbf{\Omega}_{n,k}$ is written explicitly as a logarithmic form \eqref{Omegan2log}, it can be easily compared with results in  \cite{He:2018okq}: it is sufficient to project through a fixed $Y$, which results in removing all $Y$-dependence, and to consider the differentials to act on $\lambda$. 
Finally, the volume function for $\mathrm{MHV}_n$ amplitudes is 
\begin{equation}
\Omega_{n,2} = \frac{ \langle 1\dots n\rangle^2 \left(\sum\limits_{i<j}\, [12ij] \langle Y i j \rangle\right)^2}{[\tilde Y 12]^2 \langle Y12\rangle \langle Y23\rangle \dots \langle Y1n\rangle} \,.
\end{equation}

The results for $\overline{\text{MHV}}$ amplitudes are the parity conjugate of the previous formul\ae. In particular, as for the MHV case, we do not need to triangulate the momentum amplituhedron $\mathcal{M}_{n,n-2}$ since its dimension  is already $2n-4$. The boundaries of $\mathcal{M}_{n,n-2}$ are easily found and all take the form $[\tilde Y ii+1]=0$ for $i=1,\ldots,n$. Moreover, for all points inside $\mathcal{M}_{n,n-2}$ we find $\langle Yii+1\rangle >0$ for $i=1,\ldots,n$ and $S_{i,i+1,\ldots,i+p}>0$ for $i=1,\ldots,n$ and $p=1,\ldots,n-3$.


\subsection{\texorpdfstring{NMHV$_6$}{} Amplitude}
As a next step, we consider the first example where we need to triangulate the momentum amplituhedron in order to find the volume form. The positive Grassmannian $G_+(3,6)$ is nine-dimensional, while the momentum amplituhedron $\mathcal{M}_{6,3}$ is eight-dimensional and therefore the image of the positive Grassmannian through the map $\Phi_{(\Lambda,\tilde\Lambda)}$ cannot be injective. In order to find the volume form $\mathbf{\Omega}_{6,3}$ we need to therefore focus on codimension-one cells in $G_{+}(3,6)$. There are two possible combinations of eight-dimensional cells whose images triangulate $\mathcal{M}_{6,3}$:
\begin{equation}
\mathcal{T}_1=\{(123)=0,(345)=0,(561)=0\}\,,\qquad \mathcal{T}_2=\{(234)=0,(456)=0,(612)=0\}\,,
\end{equation}
where by $(ijk)=0 $ we denote the cell in $G_{+}(3,6)$ for which the minor $(ijk)$ vanishes. 
The volume form can then be written as follows
\begin{equation}
\label{cells63}
\mathbf{\Omega}_{6,3} =\mathbf{ \Omega}_{6,3}^{(612)} +\mathbf{ \Omega}_{6,3}^{(234)} +\mathbf{ \Omega}_{6,3}^{(456)}   =\mathbf{ \Omega}_{6,3}^{(123)} +\mathbf{ \Omega}_{6,3}^{(345)} + \mathbf{\Omega}_{6,3}^{(561)} \,,
\end{equation}
where $ \mathbf{\Omega}_{6,3}^{(ijk)} $ is the pushforward of the logarithmic differential form on the cell $(ijk) = 0$. 

In the following we focus on $\mathbf{\Omega}_{6,3}^{(123)}$, the other terms can be found by cyclic shifts. 
We parametrize the cell for which $(123)=0$ using canonical coordinates and solve the relations \eqref{Y} to find
\begin{eqnarray}
\alpha_1 &=& \frac{\langle Y12\rangle}{\langle Y13 \rangle} \,, \, \alpha_2 = \frac{\langle Y23\rangle}{\langle Y13 \rangle} \,, \, \alpha_3 = \frac{[\tilde{Y}\hat{3}4]}{[\tilde{Y}\hat{1}\hat{3}]} \,,\, \alpha_4 = \frac{[\tilde{Y}64]}{[\tilde{Y}\hat{1}\hat{3}]} \\
\alpha_5 &=& \frac{[\tilde{Y}6\hat{1}]}{[\tilde{Y}\hat{1}\hat{3}]} \,,\, \alpha_6 = \frac{[\tilde{Y}4\hat{1}]}{[\tilde{Y}\hat{1}\hat{3}]} \,,\, \alpha_7 = \frac{[\tilde{Y}45]}{[\tilde{Y}64]} \,,\, \alpha_8 =  \frac{[\tilde{Y}56]}{[\tilde{Y}64]}\,,
\end{eqnarray}
where we have denoted the following shifted variables
\begin{equation}
\hat{\tilde{\Lambda}}_1 = \tilde{\Lambda}_1 + \frac{\langle Y23\rangle}{\langle Y13\rangle} \tilde{\Lambda}_2 \,,\, \hat{\tilde{\Lambda}}_3 = \tilde{\Lambda}_3 + \frac{\langle Y12\rangle}{\langle Y13\rangle} \tilde{\Lambda}_2 \,.
\end{equation}
One can notice that  the canonical variables are just an uplift of the formula found in \cite{He:2018okq}\footnote{The attentive reader will notice that we have a discrepancy of signs w.r.t. \cite{He:2018okq}. In our formul\ae~they are such that the canonical coordinates are all positive for positive data.}.
The push-forward is computed as
\begin{equation}\label{volform.n6k3}
\mathbf{\Omega}_{6,3}^{(123)} =  \bigwedge\limits_{i=1}^8 \mathrm{dlog} \alpha_i \,,
\end{equation}
which, using \eqref{volumefc}, leads to the following explicit form for the volume function 
{\small
\begin{eqnarray}\label{ans.n6k3}
\Omega_{6,3}^{(123)}&=&\frac{\left(\langle Y12 \rangle [12456] + \langle Y13\rangle [13456] + \langle Y23\rangle [23456] \right)^2 ([ \tilde Y45] \langle 12345\rangle +[ \tilde Y46] \langle 12346\rangle+ [ \tilde Y56] \langle 12356\rangle )^2}{S_{123} \langle Y 12\rangle \langle Y23\rangle  [\tilde{Y}45] [\tilde{Y}56] \langle Y1| 5+6 |4\tilde{Y}] \langle Y3| 4+5 |6\tilde{Y}] }\,.   \nonumber \\  
\end{eqnarray}
}
After using our procedure \eqref{extract} for extracting the amplitude, we find that the expression \eqref{ans.n6k3}  reduces to the formula found in \cite{He:2018okq}. While for the denominator it can be easily seen, the numerator requires a more careful analysis. The reader can convince oneself that the first  bracket in the numerator  will reduce to the ``$d^6\tilde\lambda$" part:
\begin{equation}
\left(\langle Y12 \rangle [12456] + \langle Y13\rangle [13456] + \langle Y23\rangle [23456] \right)^2\to \delta^4(q)(\tilde\eta_4 [56]_{\tilde\lambda}+\tilde\eta_5 [64]_{\tilde\lambda}+\tilde\eta_6 [45]_{\tilde\lambda})^2\,,
\end{equation} 
while the second bracket corresponds to the part proportional to ``$d^6\lambda$":
\begin{equation}
\left([ \tilde Y45] \langle 12345\rangle +[ \tilde Y46] \langle 12346\rangle+ [ \tilde Y56] \langle 12356\rangle  \right)^2 \to \delta^4(\tilde{q}) \left(\eta_1 \langle 23\rangle_{\lambda}+\eta_2 \langle 31\rangle_{\lambda}+\eta_3 \langle 12\rangle_{\lambda}\right)^2\,.
\end{equation}
Finally, we can write the complete volume form $\mathbf{\Omega}_{6,3}$ by using \eqref{volform.n6k3} and shifting labels:
\begin{equation}
\mathbf{\Omega}_{6,3}   =\mathbf{ \Omega}_{6,3}^{(123)} +\mathbf{ \Omega}_{6,3}^{(345)} + \mathbf{\Omega}_{6,3}^{(561)} =\mathbf{ \Omega}_{6,3}^{(123)} +\left.\mathbf{ \Omega}_{6,3}^{(123)} \right|_{i\to i+2}+\left.\mathbf{ \Omega}_{6,3}^{(123)} \right|_{i\to i+4} \,,
\end{equation}
and similarly for the volume function.
One can check that the spurious divergencies, appearing as poles of the type $\langle Yi| j+k |l\tilde{Y}]$ in $\mathbf{\Omega}_{6,3}^{(123)}$,  cancel in the sum and the form $\mathbf{\Omega}_{6,3}$ diverges logarithmically on the 15 boundaries of the momentum amplituhedron $\mathcal{M}_{6,3}$:
\begin{equation}
\langle Yii+1\rangle=0\,\,, i=1,\ldots,6\,,\qquad [\tilde Yii+1]=0\,, \, i=1,\ldots,6\,,\qquad S_{i,i+1,i+2}=0 \,, i=1,2,3  \,.
\end{equation}
Moreover, it is also easy to verify that for all points inside $\mathcal{M}_{6,3}$ one has $\langle Yii+1\rangle>0$ and $[\tilde Yii+1]>0$. This immediately implies that the two-particle Mandelstam variables are positive. It is however not true for the three-particle Mandelstam variables.  Let us focus first on
\begin{eqnarray}
S_{123} &=&  \langle Y 12\rangle [\tilde{Y} 12] + \langle Y 13 \rangle  [\tilde{Y} 13]  + \langle Y 23\rangle   [\tilde{Y} 23] \,.
\end{eqnarray}
It is clear that the first and last term in this expansion are explicitly positive, however the middle term has no definite sign. 
Using \eqref{Y} we can further expand the Mandelstam variable to get
\begin{eqnarray}
S_{123} = (123) &\bigg[ &\langle 1 \rangle^\perp \big[ (145) [12345] + (146) [12346] + (156) [12356] + (456) [23456] \big] \nonumber \\ &+& \nonumber \langle 2 \rangle^\perp  \big[(245) [12345] + (246) [12346] + (256)[12356] -(456)[13456 ] \big] \\ \nonumber
&+&\langle 3 \rangle^\perp \big[(345) [12345] + (346)  [12346] +(356) [12356] + (456)[12456]\big]  \bigg]\\
+ (456) &\bigg[ &\langle 4  \rangle^\perp \big[ (124) [12456] + (234) [23456] +(134) [13456] + (123) [12356] \big] \nonumber \\
&+& \langle 5\rangle^\perp \big[ (125) [12456] + (235) [23456] +(135)[13456] -(123)[12346]\big] \nonumber \\
&+& \langle 6 \rangle^\perp \big[(126)[12456] +(236) [23456] + (136) [13456] + (123) [12345] \big]  
\bigg] \,, \,\,\,\,
\end{eqnarray}
where we have organized the expansion to have all brackets explicitly positive. Then, the manifestly negative terms present in the expansion can in principle dominate over the positive terms making the Mandelstam variable negative. Let us first remark that a careful numerical analysis shows that $S_{123}$ is negative only in a very small subregion of $\mathcal{M}_{6,3}$ for generic positive data. Moreover, when the kinematic data is taken to be on the moment curve \eqref{moment.curve}, $S_{123}$  is positive.  
We can now state a sufficient condition for $S_{123}$ to be positive for points inside $\mathcal{M}_{6,3}$: we impose the constraints on the kinematics
\begin{eqnarray}\label{rel_kin_63}
\langle 1 \rangle^\perp \check{[1]} - \langle 2 \rangle^\perp \check{[2]} +\langle 3 \rangle^\perp \check{[3]}+ \langle 4 \rangle^\perp \check{[4]} -\langle 5 \rangle^\perp \check{[5]}+\langle 6 \rangle^\perp \check{[6]} > 0\,,
\end{eqnarray}
where with $[\check{i}]$ we denote the five-bracket with the index $i$ omitted. By studying the remaining independent three-particle Mandelstam variables, \emph{i.e.} $S_{234}$ and $S_{345}$, we find the same type of relations \eqref{rel_kin_63} with the signs cyclically shifted by, respectively, one and two positions. For all kinematic data satisfying these three conditions, the momentum amplituhedron $\mathcal{M}_{6,3}$ is a positive geometry.
At the moment it is  unclear what is the geometric interpretation of these inequalities.

\bigskip

\noindent {\bf Factorization properties.} One important property of the amplitudes is that they factorize into products of smaller amplitudes when planar Mandelstam variables vanish. This is reflected in the amplituhedron geometry in the fact that, when we approach one of its boundaries, then the volume form factorizes. In the ordinary amplituhedron, the statement is even more general: the geometry itself factorizes as a cartesian product of two positive geometries. For the momentum amplituhedron the situation is slightly more involved and the factorization properties rather come from the amalgamation of on-shell diagrams inside the positive Grassmannian \cite{ArkaniHamed:2012nw}.  To perform the amalgamation we  need to start with two planes $C_L \in G(k_L,n_L)$ and $C_R \in G(k_R,n_R)$, where $n_{L,R}$ denote the number of particles in the left and right diagram, respectively, and $k_{L,R}$ is their respective helicity. Then we take their direct product, which brings us to $\hat{C} \in G(k_L+k_R,n_L+n_R)$, and subsequently we project the product to $C \in G(k_L+k_R-1,n_L+n_R-2)$.
As a result, the $C$-matrix describing the cell where the factorization takes place is composed of the two overlapping $C$-matrices corresponding to the left and the right amplitude.  

To illustrate how the amalgamation works in the context of the momentum amplituhedron, we study it in details for $n=6$ and $k=3$. We encounter three different types of amalgamations, depending on which boundary we approach. Let us start by taking $S_{123}\to 0$. This boundary is parametrized by a seven-dimensional positroid cell for which $(123)=(456)=0$. This cell  can be written in terms of positive coordinates as
\begin{equation}
\left. C_{6,3}\right\vert_{S_{123}= 0} =
  \left( {\begin{array}{*{6}{c}}
   1 & \alpha_5+\alpha_7 & \alpha_5 \alpha_6 & \multicolumn{1}{c|}{0} & 0 & 0\\ \cline{3-6}
   0 & \multicolumn{1}{c|}{1} & \alpha_6 & \multicolumn{1}{c|}{\alpha_2+\alpha_4 } & \alpha_2 \alpha_3 & 0 \\ \cline{1-4} 
   0 & \multicolumn{1}{c|}{0} & 0 & 1 & \alpha_3 & \alpha_1\\
  \end{array} } \right).
\end{equation}
This matrix can be regarded as coming from the amalgamation of two positive matrices corresponding to four-point MHV amplitudes. We can indeed recognize that the two matrices inside the $(2\times 4)$ boxes are positive and both correspond to $\mathcal{M}_{4,2}$.

The second type of boundaries we consider is $[\tilde Y ii+1]\to 0$. Let us focus on $[\tilde{Y}56] \rightarrow 0$. We expect this limit to describe the case when particles $5$ and $6$ become collinear, and the amplitude $\mathcal{A}^{\tt tree}_{6,3}$  reduces to $\mathcal{A}^{\tt tree}_{5,2}$. It is indeed reflected in the form of the matrix defining this boundary. We can see it by studying the seven-dimensional cell parametrizing the boundary $[\tilde{Y}56] \rightarrow 0$:
\begin{equation}
\left. C_{6,3}\right\vert_{[\tilde{Y}56]= 0} =
  \left( {\begin{array}{*{6}{c}}
\rowcolor[gray]{0.9}   1 & \alpha_3+\alpha_5+\alpha_7 & (\alpha_3+\alpha_5) \alpha_6 & \alpha_3 \alpha_4& \multicolumn{1}{c|}{0}  &\cellcolor{white}  0\\ 
 \rowcolor[gray]{0.9}   0 & 1 & \alpha_6 & \alpha_4 & \multicolumn{1}{c|}{\alpha_2}   & \cellcolor{white} 0 \\ \cline{1-5}
      0 & 0 & 0 & 0 & 1  & \alpha_1\\ 
  \end{array} } \right),
\end{equation}
where one can recognize the positive matrix defining $\mathcal{M}_{5,2}$ in the upper left corner. Notice that the value of $k$ reduces in this limit.
Finally, we consider the limit $\langle Yii+1 \rangle\to 0$ which should correspond to a collinear limit with $k$ preserved. Indeed, the boundary with $\langle Y56 \rangle= 0$ corresponds to the following seven-dimensional cell in the positive Grassmannian:
\begin{equation}
\left. C_{6,3}\right\vert_{\langle Y56\rangle= 0} =\left(
\begin{array}{ccccc|c}
 \rowcolor[gray]{0.9} 1&\alpha_5+\alpha_7&\alpha_5\alpha_6&0&0&\cellcolor{white}0\\
\rowcolor[gray]{0.9}  0&1&\alpha_3+\alpha_6&\alpha_3\alpha_4&0&\cellcolor{white}0\\
 \rowcolor[gray]{0.9}  0&0&1&\alpha_4&\alpha_2&\cellcolor{white}\alpha_1
\end{array}
\right).
\end{equation}
The highlighted part corresponds to the positive $3\times 5$ matrix present in the definition of $\mathcal{M}_{5,3}$, as expected.
\bigskip

\noindent {\bf Comments on positive Mandelstam variables.}
We finish by commenting on the sufficient conditions for the kinematic data which guarantee positivity for all planar Mandelstam variables. We will study the case $k=3$ in details. Let us introduce the following combination of brackets relevant in this case:
\begin{equation}
\mathcal{P}_{i_1 i_2 i_3, j_1 j_2 j_3}=\langle i_1\rangle^\perp [i_2 i_3 j_1 j_2 j_3]-\langle i_2\rangle^\perp [i_1 i_3 j_1 j_2 j_3]+\langle i_3\rangle^\perp [i_1 i_2 j_1 j_2 j_3]\,.
\end{equation} 
We checked that, for any $n$ and $k=3$, the conditions guaranteeing positivity for all Mandelstam variables $S_{i,i+1,\ldots,i+p}$ read:
\begin{equation} \label{poscondsk3}
\mathcal{P}_{i_1 i_2 i_3, j_1 j_2 j_3}+\mathcal{P}_{j_1 j_2 j_3, i_1 i_2 i_3}>0 \quad i_1 i_2 i_3 \in I_{i,p}, \quad j_1 j_2 j_3 \in \bar{I}_{i,p}\,,
\end{equation}
where we defined $I_{i,p}:=\lbrace i,i+1,...,i+p \rbrace$. In particular we see that the relations \eqref{rel_kin_63} which we found in the previous section to have  $S_{123}>0$   can be written as
\begin{equation} 
\mathcal{P}_{123,456}+\mathcal{P}_{456, 123}>0\,.
\end{equation}

After a preliminary study of the Mandelstam variables for higher $k$ we have observed that a similar set of relations should be valid also in that case. However, it is unclear to us at the moment what will be the general form of such relations, whether they also provide necessary conditions and what is their geometric interpretation. 

\section{Conclusions and Outlook}

In this paper we have introduced a novel geometric object, the momentum amplituhedron $\mathcal{M}_{n,k}$, which computes tree-level scattering amplitudes in $\mathcal{N}=4$ SYM  directly in momentum space. To accomplish this, we have defined bosonized spinor helicity variables $(\Lambda_i,\tilde\Lambda_i)$, for which we imposed specific positivity conditions, {\it i.e.}~we demanded the matrices $\Lambda^\perp$ and $\tilde\Lambda$ to be positive. The image of the positive Grassmannian $G_+(k,n)$ through this positive data defines a positive geometry if additional constraints on kinematics are imposed. Then, the volume form on the momentum amplituhedron $\mathcal{M}_{n,k}$ encodes the tree-level amplitude $\mathcal{A}^{\tt{tree}}_{n,k}$. Additionally, we showed that the positive kinematics, when projected to the spinor helicity space, satisfies the conjecture formulated in \cite{He:2018okq}; in particular, it obeys the sign-flip conditions postulated there. 

As already mentioned, in the cases of MHV and $\overline{\text{MHV}}$ all planar Mandelstam variables are positive. In order for this to be true in other helicity sectors, and to guarantee the momentum amplituhedron to be a positive geometry, the positive region of the $(\Lambda,\tilde\Lambda)$-space needs to be restricted further.
The nature of such additional constraints on the kinematical data remains however unclear to us. Nevertheless, we checked algebraically up to a large number of particles $n$ that the restricted space is not empty. Extensive numerical tests showed that it is actually rather large. 
Providing necessary and sufficient conditions for positivity of planar Mandelstam variables is an open problem and is left for future work. 

Our paper opens various interesting avenues of investigation. 
 The first question  is whether a generalization of our construction to loop amplitudes is possible. There exists a natural extension of tree-level differential forms to loop integrands, as suggested in \cite{He:2018okq}. Bosonizing those formul\ae~in a similar fashion as for tree level would therefore be a step worth pursuing. Then the underlying positive geometry should bear similarities with  the ordinary loop-level amplituhedron. Perhaps the most fascinating question is  whether we can extend our construction to other theories. For instance, our work could shed light on positive geometries in twistor theories in higher dimensions \cite{Geyer:2018xgb, Geyer:2019ayz}. Moreover, the momentum amplituhedron is formulated directly in spinor helicity variables, which are universal variables for massless theories in four dimensions (and beyond). This opens the pathway for investigating positive geometries for non-planar, less- or non-supersymmetric theories. For instance, in \cite{Arkani-Hamed:2017mur} the differential forms  for Yang-Mills and non-linear sigma model were found. These forms do not have logarithmic singularities, which would indicate that there is no underlying positive geometry. However, already for $\mathcal{N}=4$ SYM one needs to factorize $\delta^4(q)$ to get a logarithmic form on the spinor helicity space, see \cite{He:2018okq}. Nevertheless, this problem disappears when we consider the forms on the momentum amplituhedron, as we showed in this paper. We anticipate that similar, but more complicated, behaviour might be possible for less- or non-supersymmetric theories.

\section{Acknowledgments}
This work was partially funded by the Deutsche Forschungsgemeinschaft (DFG, German Research Foundation) -- Projektnummern 404358295 and 404362017. M. P. would like to thank ``Fondazione A. Della Riccia'' for financial support.

\appendix
\section{Orthogonal complements}\label{App:perp}
In this appendix we set the conventions for orthogonal complements we use in the main body. Let us consider a matrix 
\begin{equation}
B=\left( \begin{matrix}
b_{11}&b_{12}&\ldots&b_{1n}\\b_{21}&b_{22}&\ldots&b_{2n}\\\vdots&\vdots&\ddots&\vdots\\b_{k1}&b_{k2}&\ldots&b_{kn}\\
\end{matrix}\right).
\end{equation}
It describes a $k$-plane $\mathcal{B}$ in an $n$-dimensional space. We can therefore define its orthogonal complement $\mathcal{B}^\perp$: an $(n-k)$-plane in $n$ dimensions. Such plane can be parametrized by an $(n-k)\times n$ matrix
\begin{equation}
B^\perp=\left( \begin{matrix}
b^\perp_{11}&b^\perp_{12}&\ldots&b^\perp_{1n}\\b^\perp_{21}&b^\perp_{22}&\ldots&b^\perp_{2n}\\\vdots&\vdots&\ddots&\vdots\\b^\perp_{n-k\,1}&b^\perp_{n-k\,2}&\ldots&b^\perp_{n-k\,n}\\
\end{matrix}\right).
\end{equation} 
Matrices related by a $GL(n-k)$ transformation acting on rows of $B^\perp$ define the same plane $\mathcal{B}^\perp$, with a different choice of basis. The maximal minors of matrices $B$ and $B^\perp$ are related to each other
\begin{equation}\label{minorsperp}
(i_1,\ldots ,i_{n-k})^\perp_B=g\, \epsilon_{i_1,\ldots,i_{n-k},j_1,\ldots,j_k} (j_1,\ldots,j_k)_B\,,
\end{equation}
where $\{i_1,\ldots,i_{n-k},j_1,\ldots,j_k\}=\{1,\ldots,n\}$ and $g$ is a scalar, independent of the minors we consider. In our considerations we will fix a particular basis of the orthogonal complement, which will fix the value for $g$. We motivate it by considering $B$ to be an element of the Grassmannian $G(k,n)$. We choose a patch in the Grassmannian such that
\begin{equation}
B=\left( \mathbb{1}_{k\times k}| b\right)\,,
\end{equation}
and the basis of its orthogonal complement to be
\begin{equation}
B^\perp=\left( -b^T | \mathbb{1}_{(n-k)\times (n-k)}\right)\,.
\end{equation}
It is easy to check, by taking $\{j_1,\ldots,j_k\}=\{1,\ldots,k\}$, that in this case
\begin{equation}
g=(-1)^{k(n-k)}\,.
\end{equation}
It is important to notice that the relation \eqref{minorsperp} is not an involution. In the most general case
\begin{equation}\label{minorsfromperp}
(j_1,\ldots,j_k)_B=\tilde{g}\, \epsilon_{j_1,\ldots,j_k,i_1,\ldots,i_{n-k}} (i_1,\ldots ,i_{n-k})^\perp_B\,.
\end{equation}
For this to agree with \eqref{minorsperp} we need to fix
\begin{equation}
\tilde{g}=(-1)^{k(n-k)}g=1\,.
\end{equation}

In order to be consistent, we need to therefore indicate which matrices from the main body of the paper we treat as $B$ and which ones as $B^\perp$. The rule we adopt is that the positive matrices will all play the role of $B$. This implies the following relations between brackets we introduced in the main text:
\begin{align}\label{round.perp}
(j_1,\ldots,j_k)&= \epsilon_{j_1,\ldots,j_k,i_1,\ldots,i_{n-k}} (i_1,\ldots ,i_{n-k})^\perp\,,\\
\langle j_1,\ldots,j_{k-2}\rangle^\perp&= \epsilon_{j_1,\ldots,j_{k-2},i_1,\ldots,i_{n-k+2}} \langle i_1,\ldots ,i_{n-k+2}\rangle\,,\\
[j_1,\ldots,j_{k+2}]&= \epsilon_{j_1,\ldots,j_{k+2},i_1,\ldots,i_{n-k-2}} [i_1,\ldots ,i_{n-k-2}]^\perp\,,
\end{align}
where the round, angle and square brackets above are the minors of the positive matrices $C$, $\Lambda^\perp$ and $\tilde\Lambda$, respectively.

\section{Proof of the relation \texorpdfstring{$(Xpq)=\langle Ypq\rangle$}{}} 
\label{App:XandY}
With the conventions for orthogonal complements from the previous appendix, we are able to prove the formula \eqref{XandY}:
\begin{align}
(X\,p\,q)&=\sum\limits_{i_1<\ldots <i_{k-2}}(i_1\ldots i_{k-2}\, p\, q)\langle i_1\ldots i_{k-2}\rangle^\perp=\\
&=\sum\limits_{j_1<\ldots<j_{n-k}}\epsilon_{i_1\ldots i_{k-2}\,p\,q\, j_1\ldots j_{n-k}} (j_1\ldots j_{n-k})^\perp \epsilon_{i_1\ldots i_{k-2}\,j_1\ldots j_{n-k}\,p\,q}\langle j_1\ldots j_{n-k}\,p\,q\rangle\\
&=\sum\limits_{j_1<\ldots<j_{n-k}} (j_1\ldots j_{n-k})^\perp\langle j_1\ldots j_{n-k}\,p\,q\rangle=\langle Y \,p\,q\rangle\,.
\end{align}

\section{Momentum Conservation}\label{App.mom.cons}

In this appendix we show that the momentum amplituhedron 
lives in the co-dimension four surface defined by the conditions:
\begin{equation}
\label{C1}
\sum_{i=1}^n\left(Y^\perp \cdot \Lambda\right)^a_i \left(\tilde Y^\perp \cdot \tilde\Lambda\right)^{\dot a}_i=0\,, \quad a, \dot{a}=1,2\,.
\end{equation} 
Let us start by observing that
\begin{equation}
0= Y^\perp \cdot Y = Y^\perp \cdot \Lambda \cdot C^\perp \,.
\end{equation}
Therefore the $2$-dimensional subspace $Y^\perp \cdot \Lambda $ is contained in the $k$-dimensional subspace $(C^\perp)^\perp=C$. Analogously, we can deduce that $\tilde{Y}^\perp \cdot \tilde{\Lambda} \subseteq C^\perp$. Then, the two subspaces $Y^\perp \cdot \Lambda$ and $\tilde{Y}^\perp \cdot \tilde{\Lambda}$ are themselves orthogonal,  as encoded in formula \eqref{C1}.

\section{Positive Mandelstam variables for $k=2$}\label{app:positiveM}
We would like to prove that, for all MHV momentum amplituhedra, every planar Mandelstam variable is positive, namely:
\begin{equation}
S_I=\sum_{j_1,j_2 \in I} \langle Y j_1 j_2 \rangle [\tilde{Y} j_1j_2 ]>0, \quad I=\lbrace i,i+1,...,i+p\rbrace \,.
\end{equation}
First,
let us  observe that, for $k=2$:
\begin{equation}
\langle Y j_1 j_2 \rangle = \langle \rangle^\perp (j_1 j_2) \,,
\end{equation}
where $\langle \rangle^\perp = \langle 1\dots n\rangle$ is positive. Then we can rewrite:
\begin{equation}
S_I=\langle 1...n \rangle \sum_{j_1<j_2 \in I, a<b} (j_1 j_2)(a b) [ab j_1 j_2 ] \,.
\end{equation}
There are only two cases for which the bracket $[ab j_1j_2]$ is negative: $a<j_1<b<j_2$ or $j_1<a<j_2<b$. In particular, we observe that if $a,b \not \in I$ then $[ab j_1j_2 ]>0$. For $b \in I$ and $a<j_1<b<j_2$, together with the term
\begin{equation}
(j_1j_2)(a b) [ab j_1j_2 ]<0\,,
\end{equation}
in the sum there are two additional terms proportional to $[ab j_1j_2 ]$:
\begin{equation}
(bj_2)(a j_1) [aj_1bj_2]+(j_1b)(a j_2) [aj_2 j_1b ]\,,
\end{equation}
both positive.
Using Schouten identity, the three terms together add up to zero:
\begin{equation}
(j_1j_2)(a b) [ab j_1j_2 ]+(bj_2)(a j_1) [aj_1bj_2]+(j_1b)(a j_2) [aj_2 j_1b ] = 0 \,.
\end{equation}
Analogously, for the case $j_1<a<j_2<b$, together with the term
\begin{equation}
(j_1j_2)(a b) [ab j_1j_2]<0\,,
\end{equation}
we have also two positive terms
\begin{equation}
(aj_2)(j_1b) [j_1b aj_2 ]+(j_1a)(j_2 b) [j_2b j_1a ] \,.
\end{equation}
Again, using Schouten identity, the three terms together add up to zero.
We therefore conclude that all negative terms are cancelled and the Mandelstam variables are positive.

\bibliographystyle{JHEP}

\bibliography{mom_ampl}

\providecommand{\href}[2]{#2}\begingroup\raggedright\begin{thebibliography}{10}

\bibitem{ArkaniHamed:2009dn}
N.~Arkani-Hamed, F.~Cachazo, C.~Cheung, and J.~Kaplan, {\it {A Duality For The
  S Matrix}},  {\em JHEP} {\bf 1003} (2010) 020,
  [\href{http://xxx.lanl.gov/abs/0907.5418}{{\tt arXiv:0907.5418}}].

\bibitem{Mason:2009qx}
L.~J. Mason and D.~Skinner, {\it {Dual Superconformal Invariance, Momentum
  Twistors and Grassmannians}},  {\em JHEP} {\bf 11} (2009) 045,
  [\href{http://xxx.lanl.gov/abs/0909.0250}{{\tt arXiv:0909.0250}}].

\bibitem{ArkaniHamed:2012nw}
N.~Arkani-Hamed, J.~L. Bourjaily, F.~Cachazo, A.~B. Goncharov, A.~Postnikov,
  and J.~Trnka, {\em {Grassmannian Geometry of Scattering Amplitudes}}.
\newblock Cambridge University Press, 2016.

\bibitem{Britto:2004ap}
R.~Britto, F.~Cachazo, and B.~Feng, {\it {New recursion relations for tree
  amplitudes of gluons}},  {\em Nucl. Phys.} {\bf B715} (2005) 499--522,
  [\href{http://xxx.lanl.gov/abs/hep-th/0412308}{{\tt hep-th/0412308}}].

\bibitem{Britto:2005fq}
R.~Britto, F.~Cachazo, B.~Feng, and E.~Witten, {\it {Direct proof of tree-level
  recursion relation in Yang-Mills theory}},  {\em Phys. Rev. Lett.} {\bf 94}
  (2005) 181602, [\href{http://xxx.lanl.gov/abs/hep-th/0501052}{{\tt
  hep-th/0501052}}].

\bibitem{Arkani-Hamed:2013jha}
N.~Arkani-Hamed and J.~Trnka, {\it {The Amplituhedron}},  {\em JHEP} {\bf 1410}
  (2014) 30, [\href{http://xxx.lanl.gov/abs/1312.2007}{{\tt arXiv:1312.2007}}].

\bibitem{Arkani-Hamed:2017tmz}
N.~Arkani-Hamed, Y.~Bai, and T.~Lam, {\it {Positive Geometries and Canonical
  Forms}},  {\em JHEP} {\bf 11} (2017) 039,
  [\href{http://xxx.lanl.gov/abs/1703.0454}{{\tt arXiv:1703.0454}}].

\bibitem{Arkani-Hamed:2017mur}
N.~Arkani-Hamed, Y.~Bai, S.~He, and G.~Yan, {\it {Scattering Forms and the
  Positive Geometry of Kinematics, Color and the Worldsheet}},  {\em JHEP} {\bf
  05} (2018) 096, [\href{http://xxx.lanl.gov/abs/1711.0910}{{\tt
  arXiv:1711.0910}}].

\bibitem{Arkani-Hamed:2017fdk}
N.~Arkani-Hamed, P.~Benincasa, and A.~Postnikov, {\it {Cosmological Polytopes
  and the Wavefunction of the Universe}},
  \href{http://xxx.lanl.gov/abs/1709.0281}{{\tt arXiv:1709.0281}}.

\bibitem{Eden:2017fow}
B.~Eden, P.~Heslop, and L.~Mason, {\it {The Correlahedron}},  {\em JHEP} {\bf
  09} (2017) 156, [\href{http://xxx.lanl.gov/abs/1701.0045}{{\tt
  arXiv:1701.0045}}].

\bibitem{Arkani-Hamed:2018ign}
N.~Arkani-Hamed, Y.-T. Huang, and S.-H. Shao, {\it {On the Positive Geometry of
  Conformal Field Theory}},  \href{http://xxx.lanl.gov/abs/1812.0773}{{\tt
  arXiv:1812.0773}}.

\bibitem{He:2018okq}
S.~He and C.~Zhang, {\it {Notes on Scattering Amplitudes as Differential
  Forms}},  {\em JHEP} {\bf 10} (2018) 054,
  [\href{http://xxx.lanl.gov/abs/1807.1105}{{\tt arXiv:1807.1105}}].

\bibitem{Witten:2003nn}
E.~Witten, {\it {Perturbative gauge theory as a string theory in twistor
  space}},  {\em Commun. Math. Phys.} {\bf 252} (2004) 189--258,
  [\href{http://xxx.lanl.gov/abs/hep-th/0312171}{{\tt hep-th/0312171}}].

\bibitem{Roiban:2004yf}
R.~Roiban, M.~Spradlin, and A.~Volovich, {\it {On the tree level S matrix of
  Yang-Mills theory}},  {\em Phys. Rev.} {\bf D70} (2004) 026009,
  [\href{http://xxx.lanl.gov/abs/hep-th/0403190}{{\tt hep-th/0403190}}].

\bibitem{Arkani-Hamed:2017vfh}
N.~Arkani-Hamed, H.~Thomas, and J.~Trnka, {\it {Unwinding the Amplituhedron in
  Binary}},  {\em JHEP} {\bf 01} (2018) 016,
  [\href{http://xxx.lanl.gov/abs/1704.0506}{{\tt arXiv:1704.0506}}].

\bibitem{Arkani-Hamed:2014dca}
N.~Arkani-Hamed, A.~Hodges, and J.~Trnka, {\it {Positive Amplitudes In The
  Amplituhedron}},  {\em JHEP} {\bf 08} (2015) 030,
  [\href{http://xxx.lanl.gov/abs/1412.8478}{{\tt arXiv:1412.8478}}].

\bibitem{Ferro:2015grk}
L.~Ferro, T.~Lukowski, A.~Orta, and M.~Parisi, {\it {Towards the Amplituhedron
  Volume}},  {\em JHEP} {\bf 03} (2016) 014,
  [\href{http://xxx.lanl.gov/abs/1512.0495}{{\tt arXiv:1512.0495}}].

\bibitem{Ferro:2018vpf}
L.~Ferro, T.~Lukowski, and M.~Parisi, {\it {Amplituhedron meets Jeffrey Kirwan
  residue}},  {\em J. Phys.} {\bf A52} (2019), no.~4 045201,
  [\href{http://xxx.lanl.gov/abs/1805.0130}{{\tt arXiv:1805.0130}}].

\bibitem{Bourjaily:2012gy}
J.~L. Bourjaily, {\it {Positroids, Plabic Graphs, and Scattering Amplitudes in
  Mathematica}},  \href{http://xxx.lanl.gov/abs/1212.6974}{{\tt
  arXiv:1212.6974}}.

\bibitem{Geyer:2018xgb}
Y.~Geyer and L.~Mason, {\it {Polarized Scattering Equations for 6D
  Superamplitudes}},  {\em Phys. Rev. Lett.} {\bf 122} (2019), no.~10 101601,
  [\href{http://xxx.lanl.gov/abs/1812.0554}{{\tt arXiv:1812.0554}}].

\bibitem{Geyer:2019ayz}
Y.~Geyer and L.~Mason, {\it {The M-theory S-matrix}},
  \href{http://xxx.lanl.gov/abs/1901.0013}{{\tt arXiv:1901.0013}}.

\end{thebibliography}\endgroup

\end{document}